# Applicability of a Novel Integer Programming Model for Wireless Sensor Networks

Alexei Barbosa de Aguiar [#1], Álvaro de M. S. Neto [#2], Plácido Rogério Pinheiro [#3], André L. V. Coelho [#4]

# Graduate Program in Applied Informatics, University of Fortaleza
Washington Soares Avenue, 1321, Room J-30, Fortaleza, CE, Brasil, 60811-905
[1] alexei@verde.com.br  [2] netosobreira@edu.unifor.br
{[3]placido, [4]acoelho}@unifor.br

*Abstract* - This paper presents an applicability analysis over a novel integer programming model devoted to optimize power consumption efficiency in heterogeneous wireless sensor networks. This model is based upon a schedule of sensor allocation plans in multiple time intervals subject to coverage and connectivity constraints. By turning off a specific set of redundant sensors in each time interval, it is possible to reduce the total energy consumption in the network and, at the same time, avoid partitioning the whole network by losing some strategic sensors too prematurely. Since the network is heterogeneous, sensors can sense different phenomena from different demand points, with different sample rates. As the problem instances grow the time spent to the execution turns impracticable.

## *I*. INTRODUCTION

Wireless sensor networks (WSNs) have been primarily used in the monitoring of several physical phenomena, such as temperature, barometric pressure, humidity, ambient light, sound volume, solar radiation, and precipitation, and therefore have been deployed in different areas of application/research, like agriculture, climate study, biology, and security.

Although they are highly useful for such applications, specially because of its low cost, WSNs offers many challenges. Their projects often demands for requirements like reliability, failure tolerance, security and long network lifetime. These implementations must be very sophisticated to overcome hard- ware extreme limitations.

The simple deployment of the approach proposed by Quintão et al. [1], while sensing different phenomena through the same WSN, can lead to inefficiency in terms of energy expenditure. With this perspective in mind, in this work, we provide an extension to the model devised by Quintão et al. [1], namely, to consider different coverage radius and sampling rates for different phenomena. We argue that the incorporation of such aspects into the model can have a significant impact on the network lifetime mainly when the spatio-temporal properties of the phenomena under observation vary a lot. The introduction of this new dimension into the model brings about novel issues to be dealt with. The critical issue relates to the concurrent routing of data related to different phenomena, as these data should be relayed to different sinks.

The rest of the paper is organized as follows. Section II presents the WSN, how do they work, the components of a sensor, the problems that can occur in a WSN and complementary knowledge to optimize the Network. Section III presents the novel integer linear programming (ILP) model for the minimization of energy expenditure in WSNs regarding the heterogeneity aspects of the sensed phenomena mentioned above. Section IV presents initial results achieved by simulation. Finally, Section V concludes the paper and comments on future work.

## II. THE WIRELESS SENSOR NETWORK

A Wireless Sensor network typically consist of a large number of small, low power, and limited-bandwidth computational devices, named sensor nodes. These nodes can frequently interact with each other, in a wireless manner, in order to relay the sensed data towards one or more processing machines (a.k.a. sinks) residing outside the network. For such a purpose, special devices, called gateways, are also employed, in order to interface the WSN with a wired, transport network. To avoid bottleneck and reliability problems, it is pertinent to make one or more of these gateways available in the same network setting, a strategy that can also reduce the length of the traffic routes across the network and consequently lower the overall energy consumption.

A typical sensor node is composed of four modules, namely the processing module, the battery, the transceiver module and the sensor module [2]. Besides the packet building processing, a dynamic routing algorithm runs over the sensor nodes in order to discover and configure in runtime the best network topology in terms of number of retransmissions and waste of energy. Due to the limited resources available to the microprocessor, most devices make use of a small operating system that supplies basic functionalities to the application program.

To supply the power necessary to the whole unit, there is a battery, whose lifetime duration depends on several aspects, among which, its storage capacity and the levels of electrical current employed in the device. The transceiver module, conversely, is a device that transmits and receives data using radio- frequency propagation as media, and typically involves two circuits, viz. the transmitter and the receiver. Due to the use of public frequency bands, other devices in the neighborhood can cause interference during sensor communication. Likewise, the operation/interaction among other sensor nodes of the same network can cause this sort of interference. So, he lower is the number of active sensors in the network, the more reliable tends to be the radio-frequency communication among these sensors. The last component, the sensor module, is responsible to gauge the phenomena of interest; the ability of concurrently collecting data pertaining to different phenomena is a property already available in some models of sensor nodes.

For each application scenario, the network designer has to consider the rate of variation for each sensed phenomenon in order to choose the best sampling rate of each sensor device. Such decision is very important to be pursued with precision as it surely has a great impact on the amount of data to be sensed and delivered, and, consequently, on the levels of energy consumed prematurely by the sensor nodes. This is the temporal aspect to be considered in the network design.

Another aspect to be considered is the spatial one. Megerian et al. [3] define coverage as a measure of the ability to detect objects within a sensor field. The lower the variation of the physical variable being measured across the area, the shorter has to be the radius of coverage for each sensor while measuring the phenomenon. This will have an influence in the number of active sensors to be employed to cover all demand points related to the given phenomenon. The fact is: the more sensors are active in a given moment, the bigger is the overall energy consumed across the net. WSNs are sometimes deployed in hostile environments, with many restrictions of access. In such cases, the network would be very unreliable and unstable if the minimum number of sensor nodes was effectively used to cover the whole area of observation. If some sensor node fails to operate, its area of coverage would be out of monitoring, preventing the correlation of data coming from this area with others coming from other areas. The localization of each sensor node is assumed to be known a priori by an embedded GPS circuit or other method [4].

A worst-case scenario occurs when we have sensor nodes as network bottlenecks, being responsible for routing all data coming from the sensor nodes in the neighborhood. In this case, a failure in such nodes could jeopardize the whole network deployment. To avoid these problems and make a robust design of the WSN, extra sensor nodes are usually employed in order to introduce some sort of redundancy. By this means, the routing topology needs to be dynamic and adaptive: When a sensor node that is routing data from other nodes fails, the routing algorithm discovers all its neighbor nodes and then the network reconfigures its own topology dynamically. One problem with this approach is that it entails unnecessary energy consumption. This is because the coverage areas of the redundant sensor nodes overlap too much, giving birth to redundant data. And these redundant data bring about extra energy consumption in retransmission nodes. The radio-frequency interference is also stronger, which can cause unnecessary retransmissions of data, increasing the levels of energy expenditure. Megerian and Potkonjak [5] present many ILP models to maximize energy consumption but not consider the dynamic time scheduling.

The solution proposed by Quintão et al. [1] is to create different schedules, each one associated with a given time interval, that activate only the minimum set of sensor nodes necessary to satisfy the coverage and connectivity constraints. The employment of different schedules prevents the premature starvation from some of the nodes, bringing about a more homogeneous level of consumption of battery across the whole network. This is because the alternation of active nodes among the schedules is often an outcome of the model as it optimizes the energy consumption of the whole network taking into account all time intervals and coverage and connectivity constraints. It is well-known that the sensing of different phenomena does not follow the same spatio-temporal profile. For instance, the temporal and spatial variations of temperature measurements in a given area can be very different from those related to humidity. Working with only one radius of coverage for all sensed phenomena entails that this radius be the smallest one. Likewise, choosing only one sampling rate for all sensed phenomena implies that this rate can keep up well with the phenomenon that varies faster.

Megerian and Potkonjak [5] present many integer linear programming models to maximize energy consumption but its paper does not consider the dynamic time scheduling. Quintão et al. [1] and Nakamura et al. [6] use a model that consider the temporal dimension as worked here but treats the phenomena as equals dimensioning by the worse of their characteristics. They use pure linear integer programming which limits the matrix sizes while the hybrid methodology presented here trespasses this barrier. Quintão et al. [7] use a genetic algorithm to solve the coverage problem, but does not address the connectivity problem, neither deal with time schedules. [8] use Lagrangean Relaxation to improve the results of previous pure linear integer programming approaches. However, this work is not time fashioned either.

### III. THE MODEL FOR OPTIMIZING ENERGY CONSUMPTION

This model was presented in Aguiar et al. [9], [10] as an extension of the work of Quintão et al. [1]. The base model had the limitations explained in section II so it was enhanced to address these gaps. New dimensions were inserted in many matrices, new constraints and auxiliary variables as well.

In order to properly model the heterogeneous WSN setting, some previous remarks are necessary:

1. A demand point is a geographical point in the region of monitoring where one or more phenomena are sensed. The distribution of such points across the area of monitoring can be regular, like a grid, but can also be random in nature. The density of such points varies according to the spatial variation of the phenomenon under observation. At least one sensor must be active in a given moment to sense each demand point. Such constraint is implemented in the model;

2. Usually, the sensors are associated with coverage areas that cannot be estimated with accuracy. To simplify the modeling, we assume plain areas without obstacles. Moreover, we assume a circular coverage area with a radius determined by the spatial variation of the sensed phenomenon. Within this area, it is assumed that all demand points can be sensed. The radio-frequency propagation in real WSNs is also irregular in nature. In the same way, we can assume a circular communication area. The radius of this circle is the maximum distance at which two sensor nodes can interact;

3. A route is a path from one sensor node to a sink possibly passing through one or more other sensor nodes by retransmission. Gateways are regarded as special sensor nodes whose role is only to interface with the sinks. Each phenomenon sensed in a node has its data associated with a route leading to a given sink, which is independent from the routes followed by the data related to other phenomena sensed in the same sensor node;

4. The energy consumption is actually the electric current drawn by a circuit in a given time period. In what follows, the elements of the novel ILP model are introduced in a step-by-step manner.

| | |
|---|---|
| $S$ | Set of sensors |
| $D$ | Set of demand points |
| $S$ | *Set* of sinks |
| $G$ | Set of phenomena (temperature, humidity, barometric pressure, etc.). Each phenomenon has its own spatio-temporal properties. The associated sampling rate has impact on data traffic, while the associated radius of coverage has impact on the number of active sensors |
| $T$ | Number scheduling periods |
| $A^d$ | Set of arcs that link sensors to demand points for phenomena |
| $A^s$ | Set of arcs that interconnects sensors |
| $A^m$ | Set of arcs that link sensors and sinks |
| $E^d(A)$ | Set of incident arcs for demand point $d \in D$ which belong to A |
| $E^s(A)$ | Set of incident arcs for sensor $s \in S$ which belong to A |
| $S^s(A)$ | Set of output arcs leaving sensor $s \in S$ which belong to A |
| $EB_i$ | Cumulated battery energy for sensor $i \in S$ |
| $EA_i$ | Energy dissipated while activating sensor $i \in S$ |
| $EM_i$ | Energy dissipated while sensor $i \in S$ is activated (effectively sensing) |
| $ET_{ij}^g$ | Energy dissipated when transmitting data from sensor *i* to sensor *j* with respect to phenomenon *g*. Such values can be different for each arc *ij* if a sensor can have its transmitter power adjusted based on the distance to the destination sensor. Each phenomenon has its own sampling rate, a parameter that impacts the total amount of data transmitted across the WSN and, consequently, the levels of energy waste |
| $ER_I$ | Energy expended in the reception of data for sensor $i \in S$ |
| $EH_J^G$ | Penalty applied when a demand point $j \in D$ for phenomenon *g* is not covered by any sensor |
| $EG_i^g$ | Penalty applied when sensor $i \in S$ is activated to unnecessarily sense the phenomenon *g* |
| $x_{ij}^{tg}$ | If sensor *i* covers demand point *j* in period *t* for phenomenon *g* |
| $z_{lij}^{tg}$ | If arc *ij* belongs to the route from sensor l to a sink in period *t* for phenomenon *g* |
| $w_l^t$ | If sensor *i* was activated in period *t* for at least one phenomenon |
| $r_i^{tg}$ | If sensor *i* was activated in period *t* for phenomenon *g* |
| $y_i^t$ | If sensor *i* is activated in period *t* |
| $h_j^{tg}$ | If demand point *j* for phenomenon *g* is not covered by any sensor in period *t* |
| $e_i$ | Energy consumed by sensor *i* considering all time periods |

The objective function (1) minimizes the total energy consumption through all time periods. The second term penalizes the existence some not covered demand points, but the solution continues feasible. It penalizes unnecessary activation for phenomenon too.

$$\min \sum_{i \in S} e_i + \sum_{t \in T} \sum_{g \in G} (\sum_{j \in D} EH_j^t h_j^{tg} + \sum_{i \in S} EG_i^{tg} r_i^{tg}) \quad (1)$$

These are the constraints adopted:

$$\sum_{ij \in E_j^d(A_g^d)} x_{ij}^{tg} + h_j^{tg} \geq 1, \forall j \in D, \forall t \in T, \forall g \in G \quad (2)$$

Constraint (2) enforces the activation of at least one sensor node i to cover the demand point j associated with phenomenon g in period t. Otherwise, the penalty variable h is set to one. This last condition will occur only in those cases when no sensor node can cover the demand point.

$$x_{ij}^{tg} \leq r_i^{tg}, \forall i \in S, \forall ij \in A_g^d, \forall t \in T, \forall g \in G \quad (3)$$

Constraint (3) turns on variable r (which means that a sensor node is actively sensing phenomenon g in period t) if its associated sensor node is indeed allocated to cover any demand point associated with g.

$$r_i^{tg} \leq y_i^t, \forall i \in S, \forall t \in T, \forall g \in G \quad (4)$$

Constraint (4) reads that sensor node i is fully active (parameter y), if it is active for at least one phenomenon of observation.

$$\sum_{ij \in E_j^s(A^s)} z_{lij}^{tg} - \sum_{jk \in S_j^s(A^s \cup A^m)} z_{ljk}^{tg} = 0, \quad (5)$$

$$\forall j \in (S \cup M - l), \forall l \in S, \forall t \in T, \forall g \in G$$

Constraint (5) relates to the connectivity issue using the flow conservation principle. This constraint enforces that an outgoing route exists from sensor node j to sensor node k if there is already an incoming route from sensor node i to sensor node j.

$$-\sum_{jk \in S_j^s(A^s \cup A^m)} z_{ljk}^{tg} = -r_l^{tg}, j = l, \quad (6)$$

$$\forall l \in S, \forall t \in T, \forall g \in G$$

Constraint (6) enforces that a route is created for phenomenon g if a sensor node is already active for that phenomenon.

$$z_{lij}^{tg} \leq y_i^t, \forall i \in S, \forall l \in (S - j), \quad (7)$$

$$\forall ij \in (A^s \cup A^M), \forall t \in T, \forall g \in G$$

In Constraint (7), if there is an outgoing route passing through sensor node i, then this sensor node has to be necessarily active.

$$z_{lij}^{tg} \leq y_i^t, \forall j \in S, \forall l \in (S - j), \quad (8)$$

$$\forall ij \in (A^s \cup A^M), \forall t \in T, \forall g \in G$$

In the same way, with Constraint (8), if there is an incoming route passing through sensor i, then this sensor has to be active.

$$\sum_{t \in T} \sum_{g \in G} (EM_i y_i^t + EA_i w_i^t) \quad (9)$$

$$+ \sum_{l \in (S-i)} \sum_{ki \in E_i^s(A^s \cup A^m)} ER_i z_{lki}^{tg}$$

$$+ \sum_{l \in S} \sum_{ij \in S_i^s(A^s \cup A^m)} ET_i^g j z_l^t ij) \leq e_i, \forall i \in S$$

The total energy consumed by a sensor node is the sum of the parcels given in Constraint (9).

$$0 \leq e_i \leq EB_i, \forall i \in S \quad (10)$$

Constraint (10) enforces that each sensor node should consume at most the energy capacity limit of its battery.

$$w_i^0 - y_i^0 \geq 0, \forall i \in S \quad (11)$$

Constraint (11) determines when the sensor node should start to sense (parameter w). If a sensor is active in the first period, its corresponding w should be set to 1.

$$w_i^t - y_i^t + y_i^{t-1} \geq 0, \forall i \in S, \forall t \in T, t > 0 \quad (12)$$

In Constraint (12), the past and current activation states of a sensor node are compared. If the sensor node was active from period t − 1 to period t, then w is set to 1.

$$x, y, z, w, h \in \{0, 1\}, e \in \mathbb{R} \quad (13)$$

## IV. COMPUTATIONAL RESULTS

In order to assess the potentialities of the novel optimization model, we have devised the simulation scenario that is described in the sequence. First of all, we have considered only one phenomenon of interest to be concurrently sensed by the same WSN. Three to six time periods were taken in consideration, although the reader should be aware that the real benefits of our extended model appear (that is, the savings in terms of energy expenditure would be more significant) when one has to deal with large numbers of time intervals.

There were 100 demand points in a square area of 10 per 10 meters. Each demand point can be assigned to either or both phenomena, but the overall coverage of each phenomenon is totally independent from each other regarding a demand point alone. In the same vein, sixteen sensor nodes were placed in the observation area. All nodes have the same processing/sensing capabilities with the possibility to sense concurrently the two phenomena. The coverage radius for the first phenomenon was set as 8.8 meters in length while the length of the coverage radius for the second phenomenon was 16 meters.

Two types of position generation were considered: Grid and random. In grid fashion, sensors and demand points are disposed regularly in columns and lines. The other scenario is created by disposing sensors and demand points in coordinates that follows a uniform probability density function inside the observation area. Due to the stochastical nature of this variation, 10 problem instances were used for each number of periods. So the results for these instances are presented as average $\pm$ standard deviation.

The sampling rate for the first and second phenomena was set as two samples per minute and one sample per minute, respectively. The length of the radius of communication between two neighbor sensors was 11 meters in size. Only one sink was placed at the middle of the regular grid. All elements of this scenario (demand points, sensors, and sink) were generated with its associated geographic coordinates. The matrix was filled with ones in those cases where the distance from the sensor and the demand point was less than or equal to the coverage radius for each phenomenon, and with zeros otherwise. Similarly, the matrices and were filled with ones in those positions where the distance between the sensor nodes or from a sensor node to the sink was less than or equal to the communication radius, and with zeros otherwise. The energy constants were calculated having as basis the values announced at a spreadsheet from a sensor node manufacturer [11]. The energy values for transmission and reception were calculated having as basis the amount of sensed data and the bit rate adopted in the devices. The penalty constant was assigned to a high value to enforce that the model covers all demand points of interest.

In order to establish a comparison, in terms of problem difficulty (variables and constraints) and energy savings (objective function values), between the heterogeneous WSN setting and its homogeneous counterpart, we have also conducted some simulations with our model considering two phenomena with the same characteristics, namely coverage radius of 8.8 meters and sampling rate of two samples per minute.

Table I shows the simulation results achieved by playing with the CPLEX platform [2] with OPL Development Studio 4.2 and Cplex 10.0. The tests were executed in Pentium Core 2 Quad 2.4 GHz 8GB of RAM memory on Windows XP. In this table, in the calculus of the real objective value we ignore the penalties and sum up only the variables. The Figures 1 to 5 provide snapshots of the scheduled plans generated for the first and second phenomena regarding the 6 time intervals considered.

In a manner as to have a better feeling of the impact of the data routing process on the energy expenditure of the WSN nodes, we have set up a second scenario with a larger area, where the length of the coverage and communication radii become smaller. By this means, there are few communication options to each sensor, and routes must be established in order to convey data to the sinks. In this new scenario, there are four sinks in the corners of the square area and our aim is to assess how many sensor nodes the model recruits to operate as routers of the traffic towards the sinks. Figure 6 the routes generated to this scenario by our model.

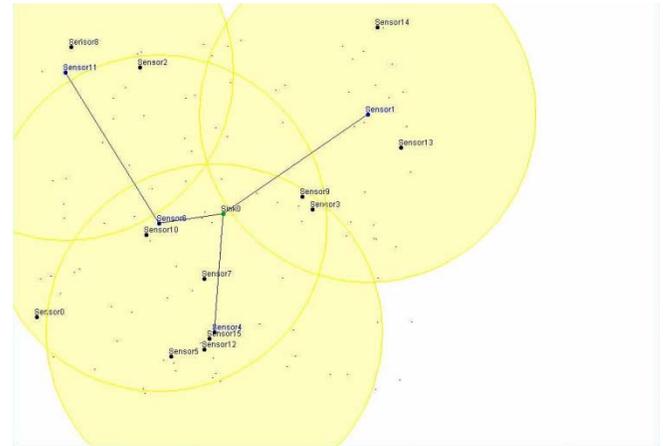

Figure 1: Phenomenon 1 - Interval 1

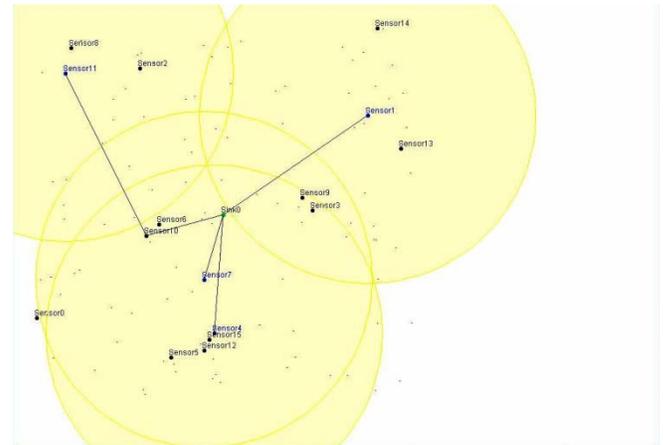

Figure 2: Phenomenon 1 - Interval 2 and 3

## V. CONCLUSION AND FUTURE WORKS

These experiments explored other possibilities of applications for this novel model. One of them is the sensors and demand point placements. Grid instances could be solved with a 0% demand point uncovered rate due to its regularity and richness of alternatives. On the other hand, random instances presented some uncovered demand point, even though they were penalized in the objective function. The reason is that some demand points are placed in some regions in the observation area that they are in the sensing coverage radius of too few sensor nodes. And the battery autonomy for each sensor node cannot supply energy for more than 3 time periods. One even more restrictive situation occurs when there is no sensor node that can reach a certain demand point. This model is prepared to handle these situations that can be found in real WSNs. This uncovered demand point penalty mechanics gives enough flexibility to deal with wider range of applications without incurring in infeasibilities.

Table I: Results for WSN problem instances

| Periods | Type | Objective | Real objective | Uncovered demand points rate | Time (s) |
|---|---|---|---|---|---|
| 1 | Grid | 3,378.62 | 3,378.62 | 0% | 1.40 |
| 2 | Grid | 6,326.80 | 6,326.80 | 0% | 2.60 |
| 3 | Grid | 9,705.62 | 9,705.62 | 0% | 44.86 |
| 4 | Grid | 12,653.80 | 12,653.80 | 0% | 56.48 |
| 5 | Grid | 19,643.00 | 19,643.00 | 0% | 26,858.81 |
| 6 | Grid | 24,017.10 | 24,017.10 | 0% | 38,885.53 |
| 1 | Random | $3,332.61 \pm 340.28$ | $4,021.61 \pm 327.24$ | $1.40 \pm 0.48$ % | $1.63 \pm 0.23$ |
| 2 | Random | $5,922.16 \pm 644.40$ | $7,159.43 \pm 457.09$ | $1.24 \pm 0.99$ % | $3.25 \pm 0.50$ |
| 3 | Random | $10,598.40 \pm 552.32$ | $8,853.31 \pm 848.73$ | $1.16 \pm 0.16$ % | $10.45 \pm 33.02$ |
| 4 | Random | $16,092.15 \pm 1,297.76$ | $12,029.01 \pm 590.71$ | $2.03 \pm 0.84$ % | $50.77 \pm 108.27$ |
| 5 | Random | $19,864.90 \pm 3,535.01$ | $14,669.99 \pm 1,517.12$ | $2.08 \pm 1.61$ % | $153.50 \pm 88.32$ |
| 6 | Random | $24,901.40 \pm 4,542.08$ | $17,385.13 \pm 1,159.51$ | $2.51 \pm 1.90$ % | $578.74 \pm 323.67$ |

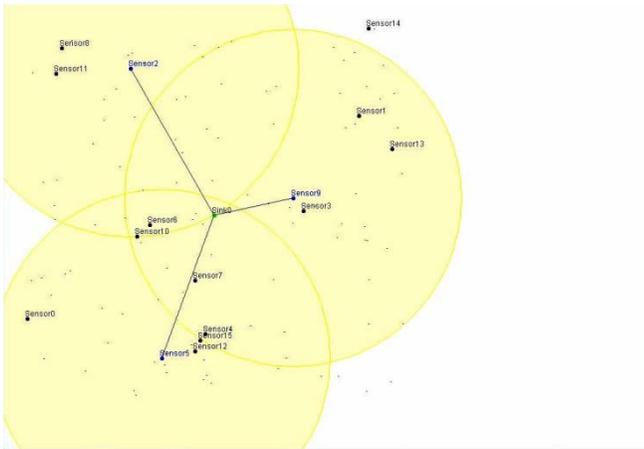

Figure 3: Phenomenon 1 - Interval 4

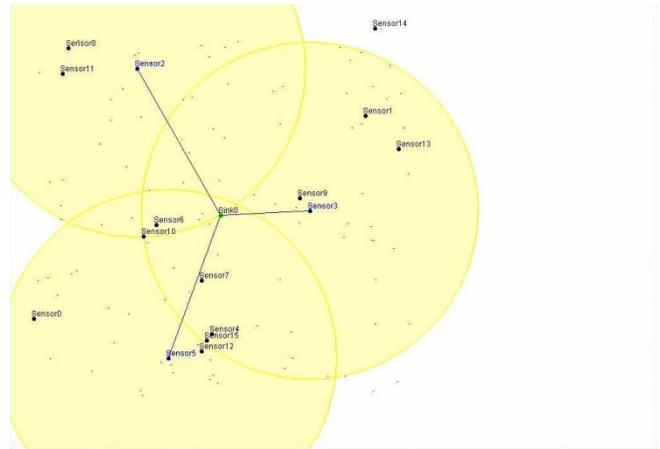

Figure 4: Phenomenon 1 - Interval 5 and 6

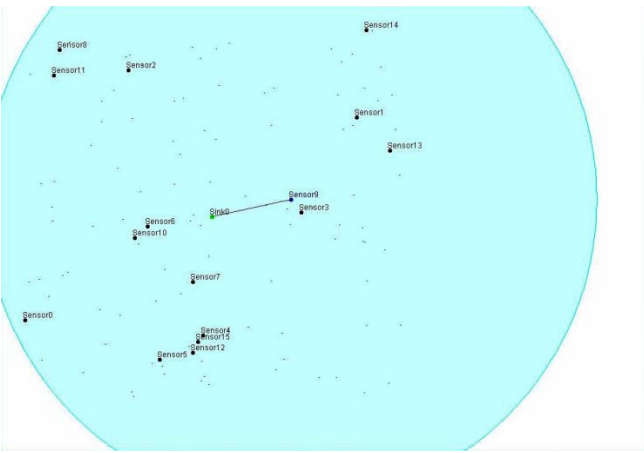

Figure 5: Phenomenon 2

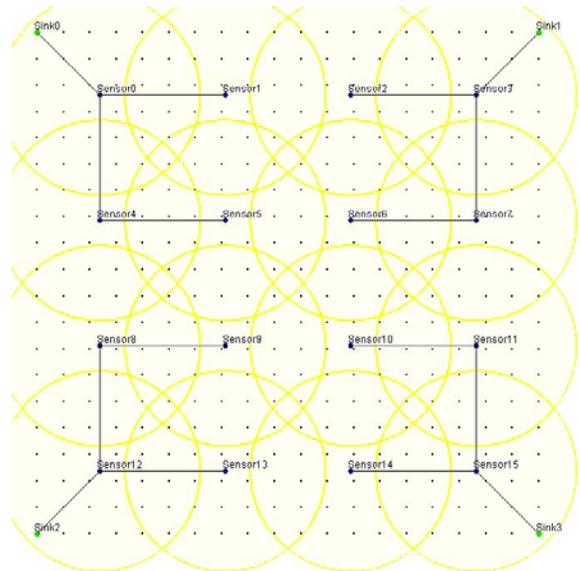

Figure 6: Routing

As combinatorial explosion quickly consumes time and memory resources, limiting WSN sizes and practical applications, the need of more sophisticated and robust methods emerges. One promising optimization area that growths and gain more attention is the hybridization of complementary approaches. In Aguiar et al. [12], an hybridization of Genetic Algorithm (GA) and ILP is used to extend the results of the homogeneous model version for optimization on WSN. In this methodology, each individual of its population generates reduced instances of the original problem. ILP is used on the solver of reduced instances and the objective value is feed backed as fitness value on that individual evaluation. This cycle evolves the solutions towards the best compromises between effectiveness and efficiency.